# Implementation of ultra-broadband optical null media via space-folding

Yichao Liu[1], Jiale Li[1], Fei Sun[1,*], Qin Liao[1], Hanchuan Chen[1], Ruihang Deng[1],

[1]*Key Lab of Advanced Transducers and Intelligent Control System, Ministry of Education and Shanxi Province, College of Physics and Optoelectronics, Taiyuan University of Technology, Taiyuan, 030024 China*

[*]sunfei@tyut.edu.cn

**Optical null medium (ONM) has garnered significant attention in electromagnetic wave manipulation. However, existing ONM implementations suffer from either narrow operational bandwidths or low efficiency. Here, we demonstrate an ultra-broadband ONM design that simultaneously addresses both challenges - achieving broad bandwidth while preserving perfect impedance matching with air for near-unity transmittance. The proposed space-folding ONM is realized by introducing precisely engineered folds into a metal channel array, creating an effective dispersion-free medium that enables independent phase control in each channel. The design incorporates optimized boundary layers implemented through gradually tapered folding structures, achieving perfect impedance matching with the surrounding medium. Beam bending effect and broadband beam focusing effect are experimentally verified using the proposed space-folding ONM. Due to its simple material requirements, broadband characteristics, and high transmittance, the proposed space-folding ONM shows potential for applications in electromagnetic camouflage, beam steering devices and ultra-compact microwave components.**

## 1. Introduction

Precision control of electromagnetic (EM) wave propagation along designed paths enables breakthrough in wireless communication systems [1,2], integrated photonics [3,4], and signature-manipulation technologies [5-11]. In essence, controlling EM wave propagation requires manipulation of phase fronts, since all wave propagation can be reduced to the interference of elementary wavelets emitted from equiphase surfaces. With the advancement of artificial structured materials, including metamaterials [12], metasurfaces [13,14], and photonic crystals[15], additional degrees of freedom have been introduced while controlling EM wave propagation. These advanced platforms now enable simultaneous manipulation of multiple wave parameters, including phase, amplitude, polarization, and dispersion. Within this landscape of artificial structured materials, optical null medium (ONM) [16,17] has merged as particularly noteworthy due to its unique capacity for perfect directional wavefront projection of EM waves. ONM is a highly anisotropic medium designed by extreme stretching from the perspective of transformation optics. ONM can guide EM propagating along its principal axis without any phase delay, thus making the output phase distribution the same as the input phase distribution. This property allows for arbitrary control of output phase by tailoring the ONM's output boundary geometry to a specific shape, as demonstrated in Fig. 1a. The ONM platform has been widely adopted for designing novel photonic devices, with recent extensions to multi-physics null media capable of simultaneous manipulation of multi-physical fields [18-21].

While ideal ONM cannot be physically realized, several practical approaches have been developed to achieve simplified ONM implementations. The primary strategies include: waveguide-structured metamaterials [16], metal channels with Fabry-Pérot resonance [22,23], subwavelength metal channels filled with gradient dielectrics [24,25], metal channels within zero-index background [26-30]. However, all current ONM implementations operate in narrow frequency bands: (i) Metamaterial-based ONMs require operation near resonant frequencies; (ii) Waveguide-structured ONMs function only near cutoff frequencies; (iii) Both empty subwavelength metal channels and dielectric-filled subwavelength metal channels are limited to Fabry-Pérot resonant frequencies; (iv) Metal channels with zero-index media exhibit narrow bandwidths, constrained by the intrinsic narrowband nature of zero-index materials. Therefore, there is still a lack of an effective method for broadband ONM implementation and experimental verification.

To overcome the inherent narrowband limitations of current ONM implementations, we propose a space-folding approach within subwavelength metal channels that corrects the equivalent optical path length in each channel. By enforcing uniform effective path lengths in channels of varying physical dimensions, our method realizes the ONM property of directional EM wave projection across an ultra-broadband frequency range (0.1-20 GHz). By adjusting two geometric parameters, i.e., the folding ratio and dilution ratio, the equivalent EM parameters of the proposed space-folding ONM can be regulated, thereby enabling the design of ONMs without modifying the shape of output boundary, as shown in Fig. 1b. Additionally, the introduction of an impedance-matching transition layer resolves the impedance mismatch between the space-folding ONM and air interface. As a proof-of-concept demonstration, we experimentally validate both beam-bending and broadband focusing effects using the space-folding ONM.

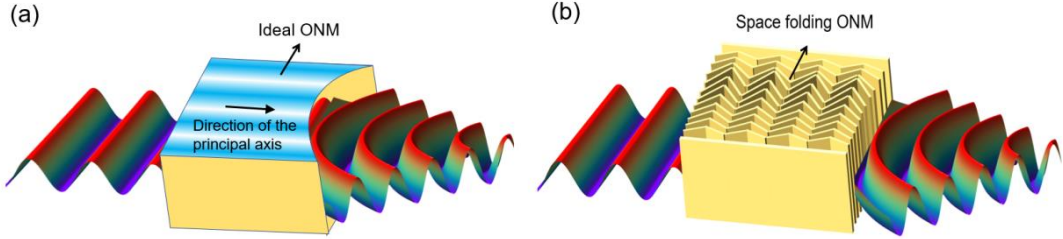

Fig. 1. Achieving the EM wave focusing effect by (a) modifying the shape of the ONM at the output boundary, (b) space-folding ONM constructed by metal channel array with folds.

## 2. Design and EM characterization of space-folding ONM

Inspired by the recent advancements in origami metamaterials [31-35], this study demonstrates that subwavelength metal array with space folding fine structures can effectively generate a space folding effect for transverse magnetic(TM)-polarized EM waves, thus realizing the space-folding ONM with adjustable output phase distributions. The structural schematic diagram of the space-folding ONM is shown in Fig. 2a. We define two parameters characterizing the folds, i.e., the folding ratio $\alpha$ and the dilution ratio $\gamma$. The folding ratio is defined as the ratio of the total length of metal channels $L$ to its lateral length $L_x$, i.e., $\alpha = L/L_x$, as shown in the upper panel of Fig. 2a. The dilution ratio is defined as the ratio of the channel width $d$ to the wavelength $\lambda$, i.e., $\gamma = d/\lambda$, as shown in the upper panel of Fig. 2a. Note the folding boundaries are not constrained to straight configurations, and- they can be flexibly designed with arbitrary curvatures, as demonstrated in the lower panel of Fig. 2a. Now we use parametric retrieval method [36,37] to illustrate how these two parameters specifically alter the effective permittivity $\varepsilon$, relative impedance $Z = \eta_1/\eta_0$ ($\eta_1$ is the impedance of the effective medium and $\eta_0$ is the impedance of air), permeability $\mu$, and refractive index $n$ of the space-folding ONM.

First, we examine the impact of the folding ratio $\alpha$ on the effective EM parameters at a fixed $\gamma$. Simulated results in Fig. 2b show that $\alpha$ equivalently changes the effective permittivity $\varepsilon$ while keeping the relative permeability $\mu$ a constant. From the perspective of geometric optics, $\alpha$ represents the effective refractive index $n$, as depicted in Fig. 2b, where the effective refractive index varies consistently with respect to $\alpha$. A higher folding ratio leads to a greater phase change for EM waves, as they travel a longer effective optical paths compared to the same lateral distance, which is clearly observed in Fig. 2c, where an increased folding ratio results in a noticeable backward shift of the EM wavefront at the output boundary.

Next, we examine the impact of the dilution ratio $\gamma$ on the effective parameters at a fixed $\alpha$. Simulated results in Fig. 2d show that increasing $\gamma$ leads to the effective permittivity $\varepsilon$ and refractive index $n$ deviate from its maximum value predicted by geometric optics. From the perspective of wave optics, an increased dilution ratio $\gamma$ indicates weakened confinement of the EM waves, because the folds exert less significant effects on areas that are distant from the fold boundaries. Therefore, a higher dilution ratio leads to a smaller phase change, which is clearly observed in Fig. 2e, where an increased dilution ratio results in a noticeable forward shift of the EM wavefront at the output boundary. The dual-parameter dependence of EM properties is systematically characterized in Figs. 2f and 2g, where Fig. 2f maps the refractive index and Fig. 2g shows the relative impedance as simultaneous functions of the folding ratio $\alpha$ and dilution ratio $\gamma$. This comprehensive visualization enables precise identification of optimal ($\alpha$, $\gamma$) pairs for target performance.

Now, the broadband performance of the designed space-folding ONM is numerically verified. The simulated material parameters $n$, $\varepsilon$, $\mu$, and $Z$ for varied frequencies are shown in Fig. 2d, which indicate that for TM-polarized EM waves, the equivalent permittivity and permeability of the space-folding ONM can be represented as: $\varepsilon_y = 2.83 + 0i$, $\varepsilon_x \rightarrow \infty$ (as EM waves cannot propagate along the $y$ direction), and $\mu_z = 1$ within an ultra-broadband frequency range (from 0.1 GHz to 20 GHz). Therefore, the space-folding ONM maintains frequency-invariant effective optical path lengths, enabling broadband directional EM wave projection across the operational spectrum.

The above results reveal a persistent impedance mismatch between the space-folding ONM and free space, leading to non-negligible boundary reflections that may compromise its performance in practical implementations. In most applications (e.g., invisibility cloaks), the ONM's impedance must match that of air at the boundaries. To achieve this, we design gradient folding ratio matching layers to eliminate the impedance mismatch between air and the central space-folding ONM (which has a fixed folding ratio), as illustrated in Fig. 2i. These matching layers enable ultrahigh transmittance across an ultra-broadband frequency range from 0.1 to 20 GHz (blue curve, Fig. 2j). By contrast, the unmatched ONM exhibits Fabry-Pérot resonances that degrade transmittance (red curve, Fig. 2j).

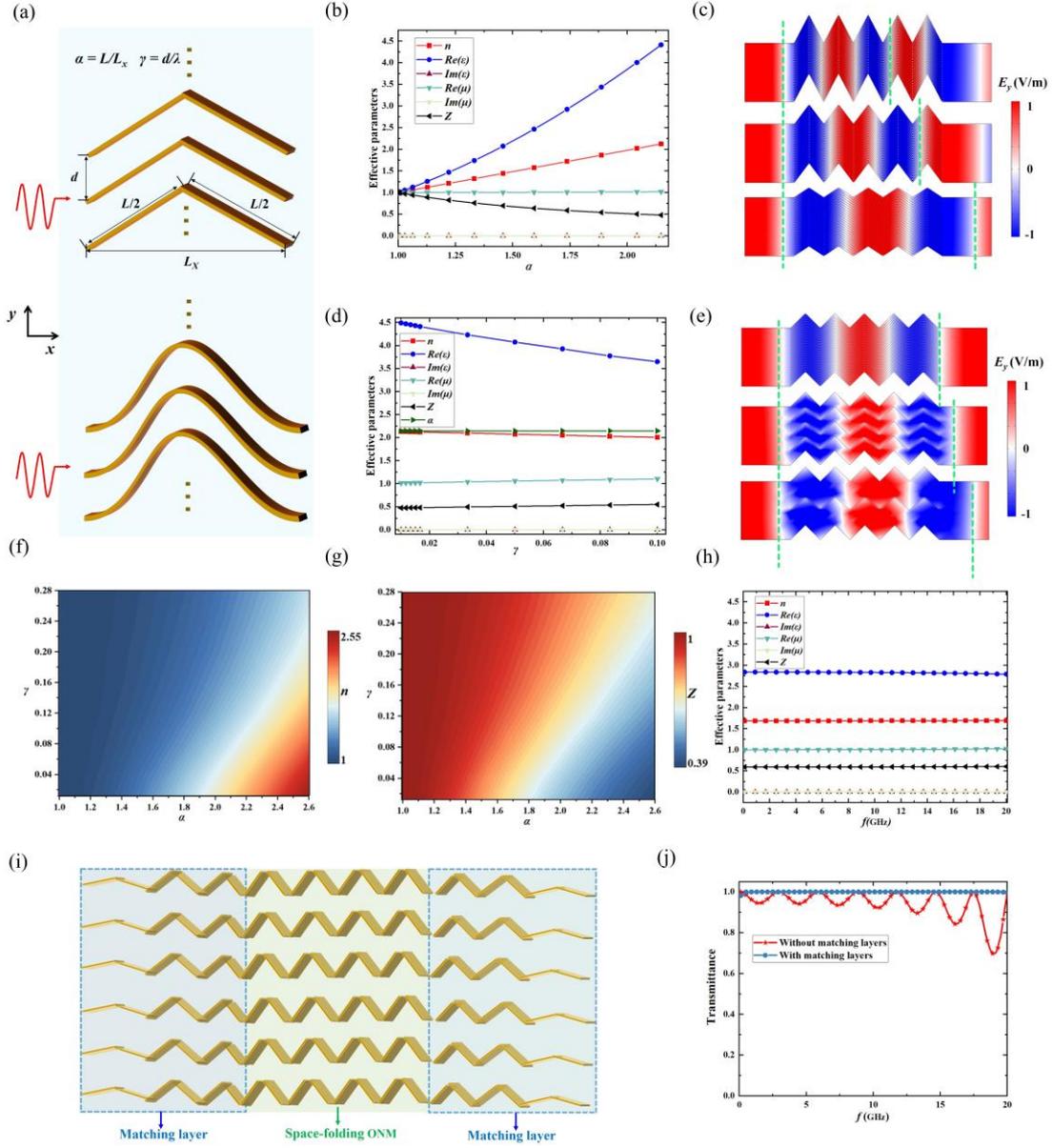

Figs. 2. (a) Structural schematic diagram of the space-folding ONM and its two key geometrical parameters, i.e., the folding ratio $\alpha$ and the dilution ratio $\gamma$. (b) The relationship between $\alpha$ and the equivalent parameters, where $\gamma = 0.01667$. (c) Electric field distribution of a plane wave passing through the space-folding ONM with fixed $\gamma = 0.01667$ and varied $\alpha$, i.e., $\alpha = 2.1767$, 1.667, and 1.2 from top to bottom. The green dashed lines represent the isophase surfaces of the EM waves before and after passing the space-folding ONM. (d) The relationship between $\gamma$ and the equivalent parameters, where $\alpha = 2.147$. (e) Electric field distribution of a plane wave passing through the space-folding ONM with fixed $\alpha = 1.414$ and varied $\gamma$, i.e., $\gamma = 0.01667$, 0.1, and 0.2 from top to bottom. (f) The variation of $n$ when $\alpha$ and $\gamma$ change simultaneously. (g) The variation of $Z$ when $\alpha$ and $\gamma$ change simultaneously. (h) The relationship between the frequency and the equivalent parameters, where $\alpha = 1.802$ and fixed $d$ ($\gamma$ ranges from 0.000334 to 0.0667). (i) Structural diagram of the space-folding ONM with matching layers. (j) Transmittance of the EM waves passing through the space-folding ONM with and without the matching layers.

## 3. Arbitrary wavefront manipulation using space-folding ONM

The space-folding ONM enables arbitrary wavefront engineering via independent control of effective EM parameters in individual channels. This precise control is achieved through judicious optimization of the folding ratio $\alpha$ and dilution ratio $\gamma$. To demonstrate this capability, we present two representative examples: (i) a beam bender, and (ii) a beam focuser.

The beam benders are designed using two types of space-folding ONM. For type-I space-folding ONM as depicted in Fig. 3a, the folding ratio increases gradually along the positive $y$ axis while maintaining a constant dilution ratio. The corresponding effective medium can be written as $\mu_1 = 1$, $\varepsilon_1 = diag(\infty, n_1^2(y), \infty)$, where $n_1(y)$ is a linear function of $y$ that can produce a linear output phase distribution. For the type-II space-folding ONM in Fig. 3d, the dilution ratio decreases gradually along the positive $y$ axis while maintaining a constant folding ratio. The corresponding effective medium can be written as $\mu_2 = 1$, $\varepsilon_2 = diag(\infty, n_2^2(y), \infty)$, where $n_2(y)$ is another linear function of $y$ that can also produce a linear output phase distribution. The simulated field patterns of a Gaussian beam passing through the two beam benders are shown in Figs. 3b and 3e, respectively. The simulated results by replacing the space-folding ONM with their corresponding effective medium are depicted in Figs. 3c and 3f. These results show EM waves exhibit an equivalent response, i.e., beam bending effect, for both the actual structures (Figs. 3b and 3e) and their corresponding effective medium (Figs. 3c and 3 f).

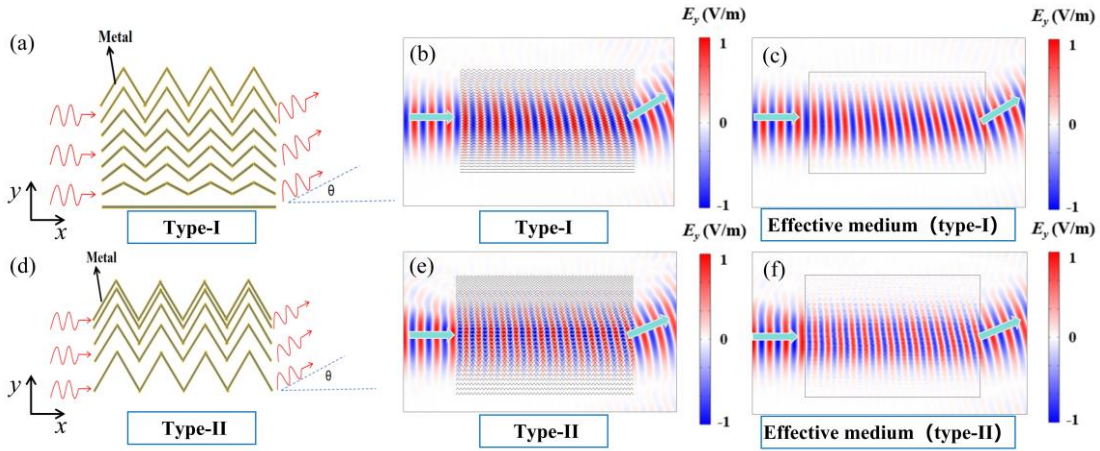

Fig. 3. (a) Schematic diagram of the beam bender with type-I space-folding ONM; (b) Simulated electric field distributions of the beam bender with type-I space-folding ONM; (c) Simulated electric field distribution for the effective medium model of the beam bender with type-I space-folding ONM, and the effective refractive index is $n_1(y) = 1 + 0.097y \cdot m^{-1}$. The wavelength is $\lambda_0 = 0.03$ m, and the Gaussian beam waist radius is $1.33\lambda_0$. (d) Schematic diagram of the beam bender with type-II space-folding ONM; (e) Simulated electric field distributions of the beam bender with type-II space-folding ONM; (f) Simulated electric field distribution for the effective medium model of the beam bender with type-II space-folding ONM, and the effective refractive index is $n_2(y) = 1.258 + 0.095y \cdot m^{-1}$. The source is a Gaussian beam with wavelength of $\lambda_0 = 0.0275$ m and beam waist radius of $1.45\lambda_0$.

As a second demonstration, we design a broadband beam focuser based on type-I space-folding ONM with matching layers. Using the two-fold symmetry of the structure, we constrain the design domain to the top-left quadrant (blue box in Fig. 5a). Within this domain, the metal channels are parametrically defined by curves, $y = y_i + A(y_i)x\sin^2(100\pi x)$, where $y_i$ represents the minimum y-coordinate of the $i$th metallic channel. This formulation yields position-dependent folding ratios $\alpha(x, y)$. Fig. 5b quantifies the resultant optical path modulation (black curve), plotting the relative optical path accumulation $\Delta L$ compared with the unfolding case, i.e., $A=0$. For effective focusing, all optical paths must satisfy the equal-phase condition, requiring specific optical path accumulations given by $\Delta L = \sqrt{F^2 + y_m^2} - \sqrt{F^2 + y^2}$ (blue line in Fig. 5b). This enables deterministic selection of $A$ values across channels to match the target $\Delta L$. Figs. 5c-d demonstrate the focusing performance under Gaussian beam illumination at distinct frequencies, exhibiting consistent beam concentration in both cases. The broadband capability is further verified in Fig. 5e, showing near-unity transmittance across a broadband, i.e., 2-9 GHz, achieved through optimized boundary impedance matching with air.

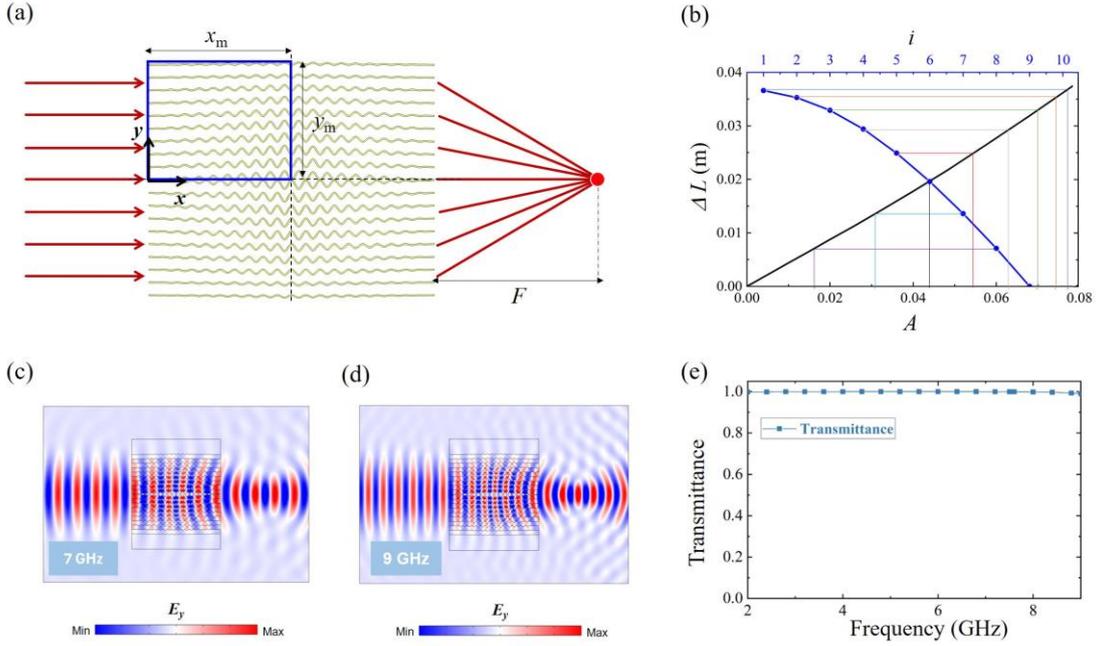

Figs. 4. (a) Schematic diagram of a beam focuser with focus length of $F$. (b) Relative optical path accumulation $\Delta L$ with different $A$ values (black) and the required $\Delta L$ at different channels with channel number $i$. Electric field distribution for a Gaussian beam incident on the beam focuser at (c) 7 GHz and (d) 9 GHz. (e) Transmittance for frequency band ranging from 2 GHz to 8 GHz.

## 4. Experiment

To experimentally verify the beam bending effect in Fig. 4, we choose copper as the metal channels and fabricate two types of space-folding ONM by tuning $α$ and $γ$ independently, as depicted in Fig. 5a. For the fabrication of the two samples, copper paper with a thickness of 0.05 mm and a height of 80 mm are folded manually to form each folded metal strip with the desired length. Subsequently, all the metal strips are fixed on a 3D-printed base to construct the two samples. The fabricated samples are shown in Figs. 5b-c, and more details about the structural parameters of the two space-folding ONM are given the Supplementary Table S1. The experiment is conducted in a microwave anechoic chamber, and a photograph of the experiment setup is shown in Fig. 5d. The fabricated sample is placed on a foam platform. A horn antenna, placed 1.5 m from the foam platform and operating at the frequency of 10 GHz, serves as the EM wave source. A displacement platform, actuated by a stepper motor and equipped with a loop receiving probe, is placed behind the sample to scan the field distribution across a 120 mm × 60 mm area. Both the horn antenna and the receiving probe are connected to the vector network analyzer (VNA, Rohde & Schwarz FSV13 9 KHz-13.6 GHz), which is controlled by computer programs.

The measurement process began by manually aligning the horn antenna toward the first sample, followed by initializing and calibrating the VNA. The displacement platform was then activated to scan the field distribution at 4 mm intervals. After completing measurements for the first sample, the same procedure was repeated for the second sample, and finally, a reference measurement was taken without any sample in place. After completing the three measurement sets, the scanned data were exported to generate the field patterns shown in Figs. 5f, 5h, and 5j, corresponding to the no-sample case, the first sample (i.e., type-I space-folding ONM), and the second sample (i.e., type-II space-folding ONM), respectively. These experimental results closely match the simulation results in Figs. 5e, 5g, and 5i. Both the Type-I and Type-II structures exhibit the beam-bending effect, deflecting incident EM waves by 30° and 20°, respectively, even without matching layers. This confirms that the phase distribution at the output surface of the space-folding ONM can be effectively modulated by adjusting its two key parameters, $α$ and $γ$. Due to the difficulty of exactly replicating the simulation conditions in the experimental environment, as well as errors in the bending of the metal thin layers, there are minor discrepancies between the experimental and simulated results.

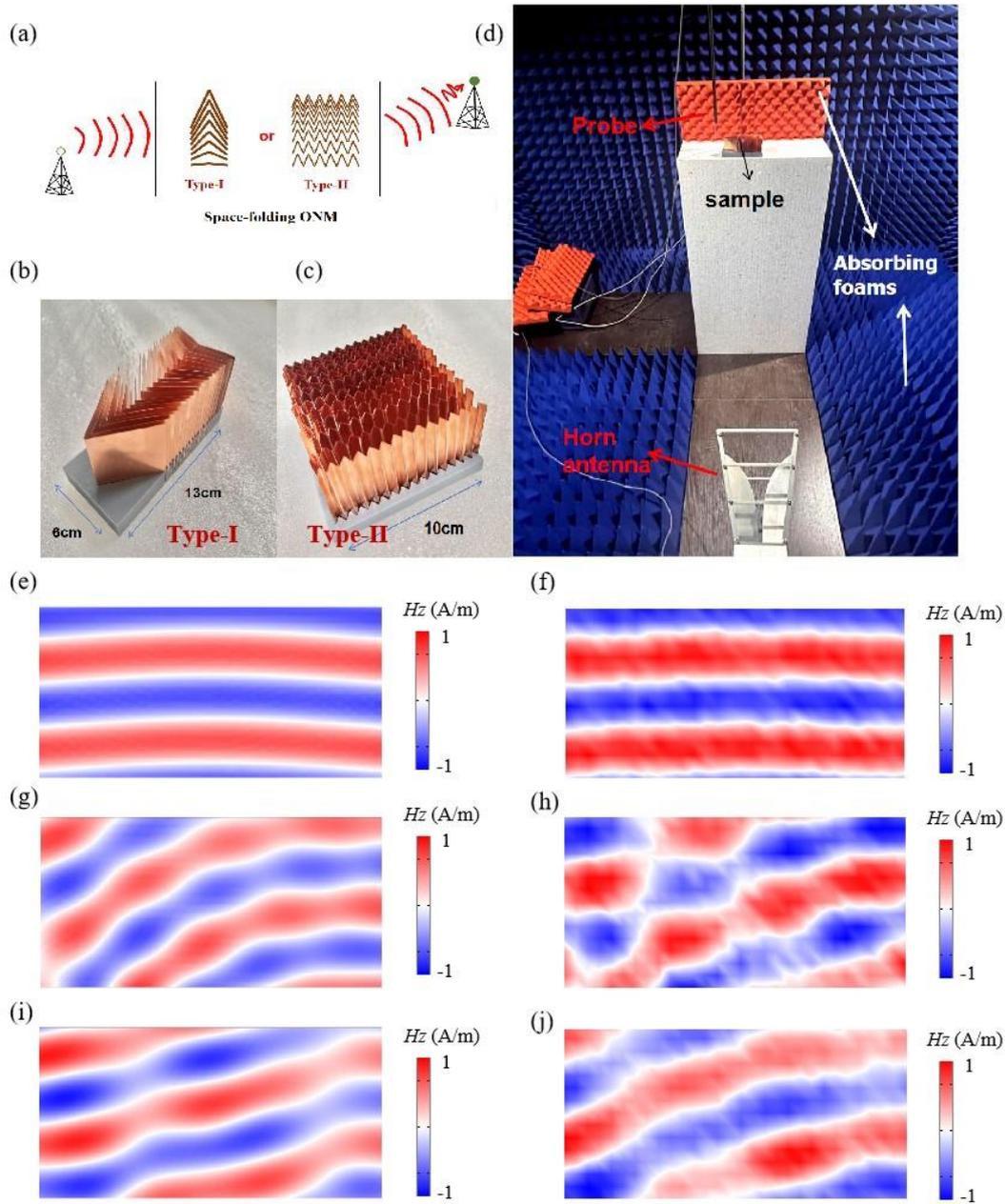

Fig. 5. (a) Schematic diagram of the beam bending experiment. (b) Photograph of the first sample: type-I space-folding ONM with gradient $\alpha$, (c) Photograph of the second sample: type-II space-folding ONM with gradient $\gamma$. (d) Photograph of the experimental setup. Simulated magnetic field distributions (e) without sample, (g) with the first sample, (i) with the second sample. Measured magnetic field distributions (f) without sample, (h) with the first sample, (j) with the second sample.

Fig. 6a presents the experimental configuration for verifying the beam focusing performance. A type-I space-folding ONM with matching layers was employed to demonstrate broadband beam focusing. The sample fabrication involved 3D printing an 80-mm-tall, 0.7-mm-thick folded thin-layer structure, which was subsequently coated with 50-μm copper tape. The process was finalized by 3D printing a base and firmly affixing the metallized structure to it. During the measurement process, we employed a stepper motor-controlled probe to scan a 200 mm linear region at a standoff distance of 5 mm from the sample surface. Reflectance was calculated by comparing electric field measurements with and without the sample, from which transmittance was derived (with negligible absorption). Fig. 6c presents the frequency-dependent transmittance curve, demonstrating consistently high transmission exceeding 95% across a frequency range from 4.6 to 8.6 GHz. To further characterize the device performance, we conducted additional scans of a 240 mm × 150 mm region behind the sample at three discrete frequencies (5 GHz, 7 GHz, and 9 GHz). The experimentally measured field distributions (Figs. 6d, 6f, and 6h) show good agreement with simulation results (Figs. 6e, 6g, and 6i), validating the structure's ultra-broadband focusing characteristics.

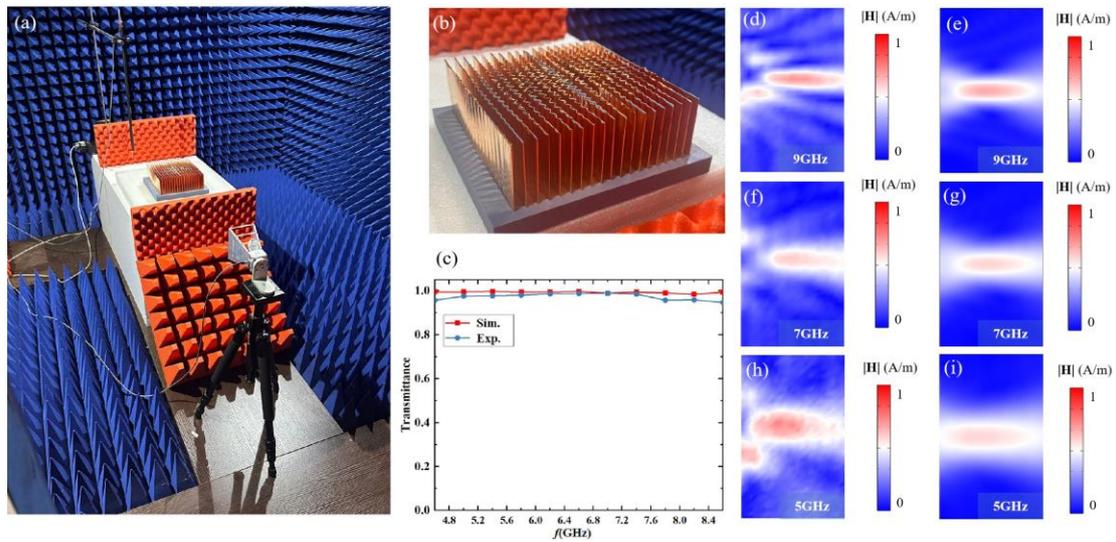

Fig. 6. (a) Photograph of the experimental setup. (b) Photograph of the beam focuser with space-folding ONM. (c) The measured transmittance (blue) and simulated transmittance (red) with varied frequency (i.e., from 4.6 GHz to 8.6 GHz) for broadband focusing. (d) Measured magnetic field pattern at 9 GHz. (e) Simulated magnetic field pattern at 9 GHz. (f) Measured magnetic field pattern at 7 GHz. (g) Simulated magnetic field pattern at 7 GHz. (h) Measured magnetic field pattern at 5 GHz. (i) Simulated magnetic field pattern at 5 GHz.

## 5. Conclusion

This study presents a simplified approach to achieving ultra-broadband ONM through the introduction of space-folding structures into subwavelength metal channel arrays. Precise control of the folding ratio ($α$) and dilution ratio ($γ$) enables independent manipulation of the effective refractive index in each channel, permitting arbitrary phase engineering at the output surface while simultaneously achieving impedance matching with air through gradual $α$ variation. Based on this design paradigm, we successfully demonstrate microwave-band beam bender and an ultra-broadband (4.6-8.6 GHz) beam focuser. Experimental results show excellent agreement with simulations, demonstrating >95% efficiency across a 61% fractional bandwidth. The proposed space-folding ONM overcomes the bandwidth limitations of conventional ONM designs and exhibits unprecedented design flexibility. This approach offers significant potential for applications in multi-band wavefront engineering, on-chip waveguiding, and broadband communication systems. Future work will focus on dynamic reconfiguration and extension to higher frequency regimes.

**Competing interests**

The authors declare no competing financial interests.

**Data availability**

The main data and models supporting the findings of this study are available within the paper. Further information is available from the corresponding authors upon reasonable request.

**Acknowledgments**

This work is supported by the National Natural Science Foundation of China (Nos. 12374277, and 12274317), and the Natural Science Foundation of Shanxi Province (202303021211054)

**Author contributions**
Y. L. and F. S. proposed the idea. J. L. performed the simulation. J. L. and Q. L. prepared the sample. J. L. and H. C. performed the measurements. Y. L. and F. S. prepared the manuscript, Y. L., J. L., F. S., Q. L., H. C., and R. D. contributed to the discussion.